# Surface plasmon dielectric waveguides

**Igor I. Smolyaninov, Yu-Ju Hung, and Christopher C. Davis**

*Department of Electrical and Computer Engineering, University of Maryland, College Park, MD 20742*

*Phone: 301-405-3255, fax: 301-314-9281, e-mail: smoly@eng.umd.edu*

**Abstract**

We demonstrate that surface plasmon polaritons can be guided by nanometer scale dielectric waveguides. In a test experiment plasmons were coupled to a curved 3 μm radius dielectric stripe, which was 200 nm wide and 138 nm thick using a parabolic surface coupler. This experiment demonstrates that using surface plasmon polaritons the scale of optoelectronic devices based on dielectric waveguides can be shrunk by at least an order of magnitude.

Dielectric waveguides form the backbone of present-day optoelectronics industry. However, in future applications, which require further miniaturization of optoelectronic components down to nanometer scales, they are going to meet their fundamental limitation set by diffraction. A usual dielectric waveguide cannot restrict the spatial localization of optical energy beyond the $\lambda_0/2n$ limit, where $\lambda_0$ is the free-space photon wavelength and $n$ is the refractive index of the waveguide. Optical waveguides based on surface plasmon polaritons (SPP) [1] offer an alternative, since at a given frequency the wavelength of SPP may be much smaller than the photon wavelength (or in other words, the effective 2D refractive index of the waveguide may be very large, as seen by the SPPs). SPP waveguides based on gaps in photonic crystal structures [2], metal stripes [3,4], or linear chains of metal nanoparticles [5,6] have been demonstrated very recently. However, all the SPP waveguides described in the literature have some potential drawbacks. For example, waveguides based on chains of nanoparticles and photonic crystal effects are relatively challenging to fabricate, while SPP waveguides based on metal stripes may suffer from edge scattering and higher losses near the SP resonance frequency where the SPP wavelength is short. Thus, alternative designs of SPP waveguides may be useful. As has been shown very recently in our SPP microscopy experiments [7,8], the diffraction limit of dielectric plasmon-optical elements may be pushed down to a scale of a few tens of nanometers (and may be even further if dielectrics with gains are used), which means that the usual $\lambda_0/2n$ diffraction limit on the performance of a dielectric waveguide can be broken. In this paper we demonstrate that surface plasmon polaritons can indeed be guided in the metal film plane by nanometer scale dielectric waveguides. In a test experiment plasmons were coupled to a curved 3 µm radius dielectric stripe, which was 200 nm wide and 138 nm thick using a parabolic surface coupler. These experiments demonstrates that using surface plasmon polaritons the scale of optoelectronic devices based on dielectric waveguides can be shrunk by at least an order of magnitude.

In our experiments 50 nm thick gold films were sputtered onto a glass substrate using a Magnetron Sputtering Machine. An overlay of PMMA film was then spin-coated and patterned using E-beam lithography. The dielectric PMMA film was about 100-200 nm thick. An example of a patterned bi-grating is shown in Fig. 1(a). It consists of individual PMMA dots on gold film surface. The grating period is 500 nm in both directions. The gold films under the PMMA layer were still intact after the gratings were developed using MIBK/IPA developer. With 502 nm laser light illumination the area of the gold film above which the PMMA grating has been formed exhibited the effect of extraordinary optical transmission similar to the one described in ref. [9]. This is illustrated in Fig.1(b), in which the area of the gold film under the PMMA grating appears similar in brightness to the areas of the gold film punctured with a periodic array of nanoholes (compare Fig.1 (b) with Fig. 4 from ref.[7] ). According to refs.[7,9,10] such periodic arrays of bumps and holes act as very efficient sources of SPPs.

In the next set of experiments we have shown that the SPPs emitted by the dielectric bi-gratings descried above may be efficiently focused by parabolically-shaped PMMA mirrors (Fig.2). Ideologically, these two-dimensional dielectric focusing mirrors are similar to the parabolic glycerin droplets, which were shown to focus the SPP field in our earlier microscopy experiments [7,8] . However, fabrication of such mirrors lithographically using PMMA is far more advantageous for real technological applications due to much higher degree of the shape control in lithographical experiments. Fig.2 illustrates that the SPP field generated over the rectangular PMMA bi-grating (similar to the one shown in Fig.1) is focused by the parabola-shaped PMMA mirror in the area between the apex and the focus point of the parabola. Comparison of Figs. 2 (a) and (b) indicates that this effect may be described by geometrical optics (ray tracing) similar to the description of SPP microscopy in refs.[7,8]. The propagation of SPP energy towards the apex of the parabolic mirror may be visualized similar to the visualization of light propagation through a fiber. Due to increased Raleigh scattering at higher laser field intensity the SPP propagation through the PMMA mirror is accompanied by SPP scattering into photons, which is clearly visible in Fig.2(c).

Finally, we were able to use the observed SPP field focusing by the parabolic PMMA mirrors in order to couple SPP field into the dielectric waveguides formed near the parabola apex. Two examples of the dielectric SPP waveguides are shown in Fig.3. According to AFM measurements (Fig.3(d)) the curved PMMA waveguide formed near the apex of the parabolic SPP coupler has 3 μm radius, 200 nm width and 138 nm thickness. Such a thin dielectric waveguide can only support a SPP guided mode (since the waveguide thickness multiplied by the PMMA refractive index is $0.4\lambda<\lambda/2$ in this experiment). On the other hand, rather efficient guiding of SPP energy down the dielectric waveguides is quite apparent from Figs 3 (b) and (c): the brightness of scattered light produced by the SPP scattering at the edge of the parabolic PMMA mirror is clearly of the same order of magnitude as the brightness of the scattered light at the output ends of the SPP waveguides. Guiding of optical energy down such thin waveguides indicates that at least an order of magnitude compactification of dielectric waveguide devices is possible if SPPs are used. In fact, according to refs.[7,8] the effective 2D refractive index of the dielectric waveguide at the frequency of surface plasmon resonance [1] diverges as seen by the plasmon polaritons, and guiding of the SPP energy over the dielectric waveguides as thin as a few tens of nanometers is possible. Compared to other SPP waveguides operating in the short-wavelength limit, the dielectric waveguides have lower losses (due to lower frequency of the surface plasmon resonance, and hence smaller imaginary part of the dielectric constant of the metal) and are easier to fabricate than the SPP waveguides based on the chains of metal nanoparticles. The lithographically created dielectric couplers and waveguides described in this paper may form the basis of future nanophotonic devices.

In conclusion, we have shown experimentally that surface plasmon polaritons can be guided by nanometer scale dielectric waveguides. Our observations demonstrate that using surface plasmon polaritons the scale of optoelectronic devices based on dielectric waveguides can be shrunk by at least an order of magnitude.

This work was supported in part by NSF grants ECS-0304046 and CCF-0508213.

# Figure Captions

**Figure 1.** (a) Electron microscope image of a PMMA bi-grating on gold film surface with approximately 0.5 µm period. The grating consists of individual PMMA dots. (b) Illuminated by 502 nm laser light, the area of the gold film above the grating exhibits the effect of extraordinary transmission and generates SPPs.

**Figure 2.** (a) Parabola-shaped PMMA mirror with a rectangular bi-grating area formed inside. The grey trapezoid near the focus point of the parabola is the image of the rectangular bi-grating obtained using geometrical optics (ray tracing). (b) After the bi-grating illumination with 502 nm laser light, the field enhancement is observed between the apex and the focus of the parabola. (c) Optical energy propagation towards the apex may be visualized at the increased laser power due to Raleigh scattering of the SPP field in the PMMA film.

**Figure 3.** (a) Linear dielectric waveguide is formed near the apex of the parabolic PMMA mirror. In (b) coupling and guiding of the optical energy down the end of the linear waveguide shown in (a) is demonstrated. (c) In a similar experiment optical energy is coupled and guided down the curved dielectric waveguide. (d) AFM image of the curved waveguide formed at the apex of the parabolic mirror. The thin horizontal stripe of PMMA is used to scatter the SPPs that may propagate through the mirror boundary.

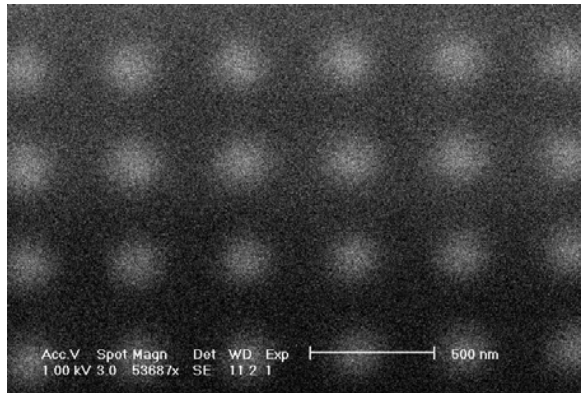

( a )

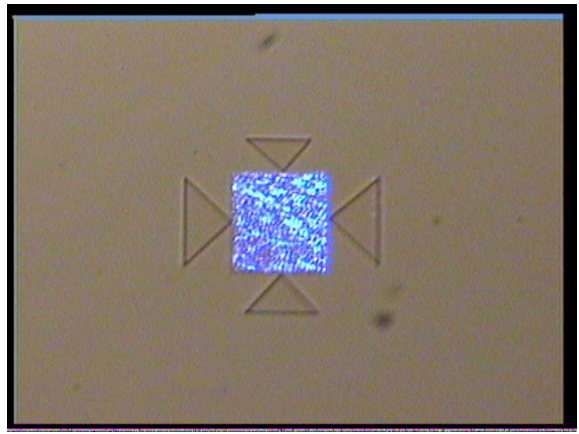

( b )

Figure 1.

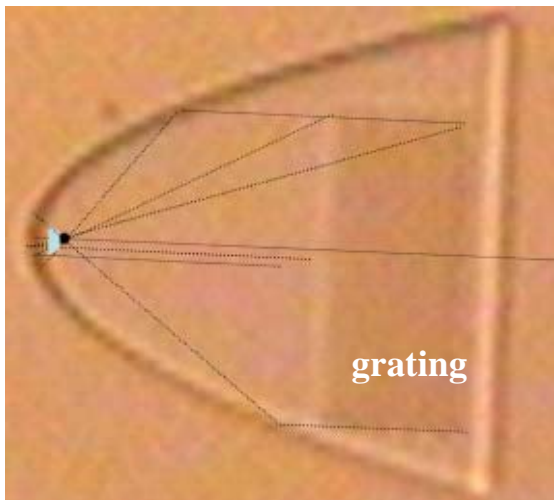

( a )

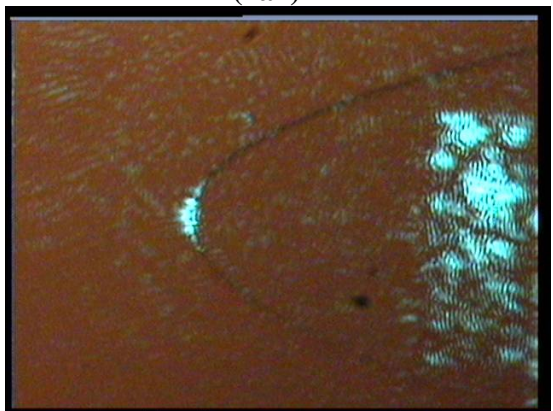

( b )

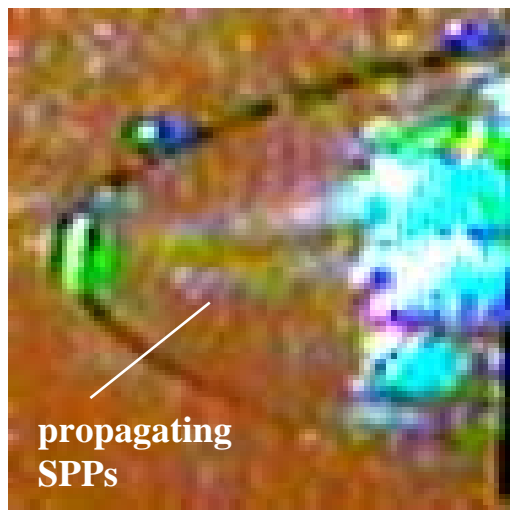

( c )

Figure 2.

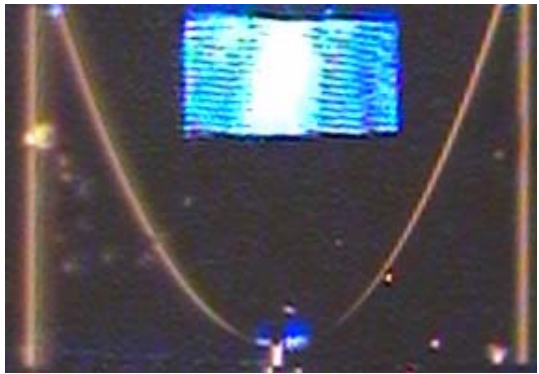

(a)

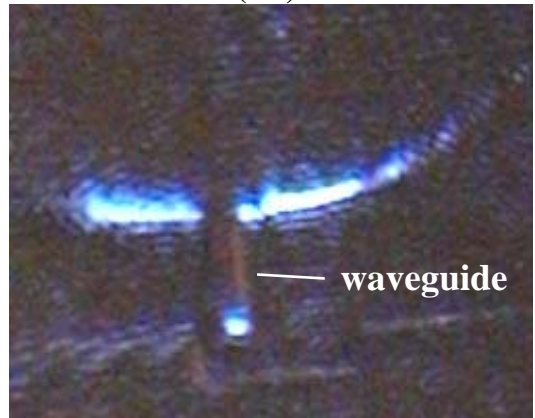

waveguide

(b)

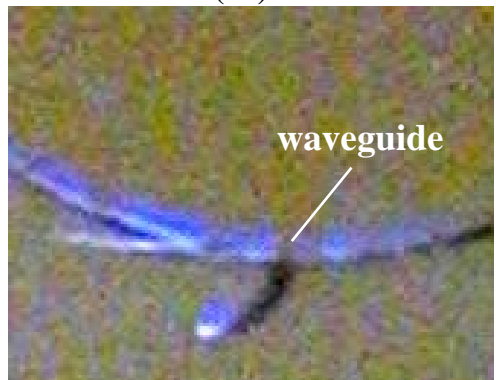

waveguide

(c)

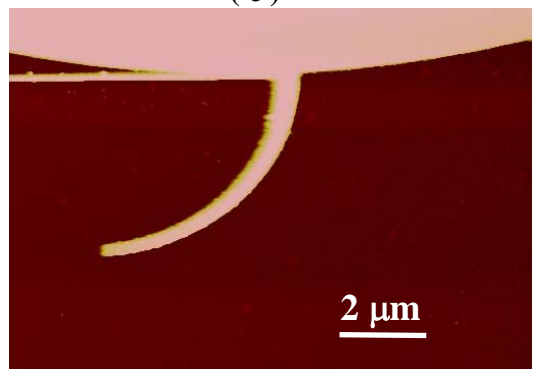

2 μm

(d)

Figure 3.